\documentclass[12pt,preprint]{aastex}

\usepackage{ragged2e}
\justifying\let\raggedright\justifying

\usepackage{threeparttable}
\usepackage{color}

\usepackage{graphicx}
\usepackage{epsfig}
\usepackage{bm}
\usepackage{subfigure}
\usepackage{multicol}
\usepackage{multirow}
\usepackage{array}
\newcommand{\PreserveBackslash}[1]{\let\temp=\\#1\let\\=\temp}
\newcolumntype{C}[1]{>{\PreserveBackslash\centering}p{#1}}
\newcolumntype{R}[1]{>{\PreserveBackslash\raggedleft}p{#1}}
\newcolumntype{L}[1]{>{\PreserveBackslash\raggedright}p{#1}}

\shortauthors{Song et. al} \shorttitle{Circular-ribbon white-light flares}

\begin{document}

\title{Investigation of white-light emission in circular-ribbon flares}

\author{Yongliang Song\altaffilmark{1,2}, Hui Tian\altaffilmark{1}}

\altaffiltext{1}{School of Earth and Space Sciences, Peking University, Beijing 100871, China; \\ huitian@pku.edu.cn}
\altaffiltext{2}{CAS Key Laboratory of Solar Activity, National Astronomical Observatories, Beijing 100012, China}

\begin{abstract}

Using observations by the \textit{Solar Dynamics Observatory} from June 2010 to December 2017, we have performed the first statistical investigation of circular-ribbon flares (CFs) and examined the white-light emission in these CFs. We find 90 CFs occurring in 36 active regions (ARs), including 8 X-class, 34 M-class, 48 C- and B-class flares. The occurrence rate of white-light flares (WLFs) is 100\% (8/8) for X-class CFs, $\sim$62\% (21/34) for M-class CFs, and $\sim$8\% (4/48) for C- and B-class CFs. Sometimes we observe several CFs in a single AR, and nearly all of them are WLFs. Compared to normal CFs, CFs with white-light enhancement tend to have a shorter duration, smaller size, stronger electric current and more complicated magnetic field. We find that for X-class WLFs, the white-light enhancement is positively correlated with the flare class, implying that the white-light enhancement is largely determined by the amount of released energy. However, there is no such correlation for M- and C-class WLFs, suggesting that other factors such as the time scale, spatial scale and magnetic field complexity may play important roles in the generation of white-light emission if the released energy is not high enough.

\end{abstract}



\section{Introduction}

Flares associated with a sudden enhancement of optical continuum emission are defined as white-light flares (WLFs).  Since the first observation of WLF in human history by Carrington (1859) and Hodgson (1859), the number of WLFs recorded in literature is very small compared to the total number of solar flares (Fang et al. 2013). Though rarely observed, WLFs are important for flare research. First, they can help us to understand the energetics of stellar flares, which are often observed in white light (WL). Second, they are usually thought to result from some extreme conditions, which greatly challenges theories of flare energy transportation (Neidig 1989) and heating mechanisms in the lower atmosphere (Ding et al. 1999a).
 
Though WLFs were discovered more than 150 years ago, a satisfactory theoretical understanding of WLFs is still lacking (Hudson 2016). Many questions remain unknown or controversial. Most observations appear to suggest that WLFs are associated with energetic events such as X-class flares. However, later it was found that some small C-class flares can also be WLFs (Matthews et al. 2003; Hudson et al. 2006; Jess et al. 2008). Many studies reveal a good correlation of the WL emission with hard X-ray (HXR) and radio emissions both in time and space (Hudson et al. 1992; Metcalf et al. 2003; Chen \& Ding 2005, 2006; Krucker et al. 2011; Hao et al. 2012; Cheng et al. 2015; Yurchyshyn et al. 2017; Song et al 2018a). These WLFs are usually believed to result from accelerated high-energy non-thermal electrons (Neidig 1989; Fletcher et al. 2008;  Kuhar et al. 2016). However, in other cases no such correlation has been found (Ryan et al. 1983; Ding et al.1994; Sylwester 2000), and it is unclear what heating mechanisms of this type of WLFs are. Theoretically, several heating mechanisms have been proposed to explain the WL emission, including electron beam bombardment (Hudson \& Ohki 1972; Aboudarham \& Hénoux 1986), Alfv\'en wave dissipation (Fletcher \& Hudson 2008), backwarming (Machado et al. 1989; Metcalf et al. 1990; Heinzel \& Kleint 2014) and chromospheric condensation (Gan \& Mauas 1994; Kowalski et al. 2015). All of them can describe or explain some characterstics of WLFs. However, due to observational limitations, it is very difficult to tell which mechanism dominates in a WLF. In addition, current observations often cannot tell whether the WL emission comes from the photosphere (Mart\'inez Oliveros et al. 2012; Watanabe et al. 2013), chromosphere (Battaglia \& Kontar 2012; Krucker et al. 2015), or both.
 
It is widely believed that the magnetic field topology plays a significant role in solar eruptions (e.g., Zhang \& Low 2005; Zhang \& Flyer 2008; Sun et al. 2015; Guo et al. 2017; Yang et al. 2017). The evolution of magnetic field structures results in the magnetic energy transportation from the solar interior to the surface, and the storage in the corona (e.g., Berger 1988; Zhang \& Low 2005). Many studies reveal that the productivity of flares or CMEs in an AR is closely related to the complexity of the magnetic field (e.g., Zirin \& Liggett 1987; Wang et al. 1991; Liu et al. 2016). Magnetic field evolution, such as flux emergence and cancelation (e.g., Zirin \& Wang 1993; Schmieder et al. 1997; Chen \& Shibata 2000; Zhang et al. 2001; Louis et al. 2015), rapid shearing motion around the polarity inversion lines (PILs; e.g., Meunier \& Kosovichev 2003; Yang et al. 2004; Shimizu et al. 2014), and sunspot rotation (e.g., Zhang et al. 2007; Yan et al. 2009, 2015, 2016; Vemareddy et al. 2016), is crucial for the occurrence of flares or CMEs. However, the role of magnetic field structure and its evolution in the production of WLFs is still poorly understood. Recently, Song et al. (2018b) identified 20 WLFs in NOAA AR 11515 during its passage across the solar disk. They found that all these WLFs occurred along a narrow ribbon-like negative flux region, with positive fluxes on both sides. In their another work (Song et al. 2018a), they found that the WL enhancement appeared at the footpoints of an erupting filament. These results suggest a close relationship between WLFs and the magnetic field environment.

The shapes of flare ribbons depend largely on the magnetic field structures of the hosting ARs. Flares sharing a similar shape of ribbon are usually associated with a similar magnetic field configuration. With recent high-resolution observations from both space-born and ground-based telescopes, several circular-ribbon flares (CFs) have been reported (Masson et al. 2009; Wang \& Liu 2012; Sun et al. 2013; Deng et al. 2013; Zhang et al. 2016; Yang et al.2015; Xu et al. 2017; Li et al. 2018). These flares are believed to be associated with a fan-spine magnetic field configuration (Sun et al. 2013). The line-of-sight photospheric magnetograms usually reveal a central unipolar-field region encompassed by the opposite-polarity magnetic field. Recently, Hao et al. (2017) and Song et al. (2018a) have identified WL emission in this type of flares. Similar to Kane et al. (1985) and Matthews et al. (2003), Hao et al. (2017) also found two types of WL kernels: an impulsive one and a gradual one. They are likely associated with different heating mechanisms. Through magnetic field extrapolation, Song et al. (2018a) found a standard fan-spine topology of a CF. They also found a flux rope under the dome-like magnetic field structure, which might be related to the production of the WLF. 

In this paper, we present results from the first statistical investigation of CFs and examine the occurrence of WL emission in these flares. We describe the observations in Section 2. Results and discussion are presented in Section 3. In Section 4 we give a brief summary.


\section{Observations}

We mainly use the data obtained by the Helioseismic and Magnetic Imager (HMI; Scherrer et al. 2012) and the Atmospheric Imaging Assembly (AIA; Lemen et al. 2012) on board the \textit{Solar Dynamics Observatory (SDO)}. HMI observes the full solar disk using the spectral line of Fe {\footnotesize I} 6173 \AA\  with a $4096 \times 4096$ CCD detector. The spatial resolution is about $1^{\prime\prime}$. The AIA instrument obtains full-disk images of the Sun, with a cadence of 12 s and 24 s in several extreme ultraviolet (EUV) and ultraviolet (UV) passbands, respectively. The spatial resolution of an AIA image is about $1.5^{\prime\prime}$.

We first searched for candidates of CFs from the \textit{SDO}/AIA 1600 \AA\ daily movies (15-min cadence) between 2010 June 1st and 2017 December 31st on the \textit{SDO} website (https://sdo.gsfc.nasa.gov/data/aiahmi/), then identified CFs from these candidates through a visual inspection of the high-cadence movies using the JHelioviewer software (M\"uller et al. 2017). In total 90 CFs from 36 ARs have been identified, including 8 X-class flares, 34 M-class flares, 47 C-class flares and 1 B-class flare. Table 1 lists these CFs. To the best of our knowledge, this is the largest sample of CFs so far. It should be noted that the ribbons of most identified CFs are nearly closed circles in the AIA 1600 \AA\  images. However, the ribbons of several identified CFs are not closed circles, but are more like half circles in the AIA 1600 \AA\ channel. These flares are marked as quasi-circular (QC) events in Table 1. The longitudinal component of the photospheric magnetic field measured by HMI reveals a pattern of one polarity surrounded by the opposite-polarity magnetic fluxes for these CFs. We have also analyzed the hard X-ray data observed by the \textit{Reuven Ramaty High Energy Solar Spectroscopic Imager} (\textit{RHESSI}; Lin et al. 2002). \textit{RHESSI} observations are available for 50 CFs. 

WLFs are very rare among low-class (C-class and below) flares. One reason may be that the WL enhancements in these flares are very weak compared to the background continuum intensity. To enhance the visibility of WL emission, we followed the method described in Song et al. (2018b) to construct the pseudo-intensity images by magnifying the difference between the HMI continuum images taken at two adjacent times. It should be noted that this method is just used to examine whether there is an impulsive enhancement of HMI continuum intensity during a flare. We find 33 WLFs from these 90 CFs, including 8 X-class, 21 M-class and 4 C-class flares. In Table 1 the events marked with the star sign are WLFs. 

\section{Results and discussion}

\subsection{Occurrence rate of WLFs in CFs}

The hosting AR numbers and \textit{GOES} classes of the 90 CFs are shown in Figure 1. The red and green dots refer to the WLFs and non-WL flares, respectively. Some ARs are very productive, each producing several CFs. For instance, NOAA AR 11476 has produced 5 CFs, and AR 11890 has produced 8 CFs. Most CFs are found in 18 ARs, each of which has produced two or more CFs. This probably suggests that in a certain period these ARs have a relatively stable magnetic field structure. The magnetic flux possibly changes after the occurrence of the previous CF. However, the fan-spine topology likely remains till the occurrence of the next CF. In contrast, each of the other 18 ARs only has one single CF, possibly suggesting destruction of the fan-spine topology by efficient reconnection during the occurrence of the CF. For the ARs with several CFs, 10 of them have produced at least one WLF. However, for the ARs with only one CF, only four of them show WL enhancement. 

For a long time, WLFs have been rarely reported. However, with high-resolution space observations, WLFs are more and more commonly observed. For instance, Watanabe et al. (2017) identified approximately 100 flares with M and X classes from observations of the Solar Optical Telescope (SOT; Tsuneta et al. 2008) on board Hinode, and found that about half of them are WLFs. Using HMI observations, Song et al. (2018b) identified 20 WLFs out of a total of 70 flares above C class (28.6\%) in NOAA AR 11515. Our observation reveals an even higher occurrence rate of WLFs in CFs. From Figure 1, we see that all the X-class CFs are WLFs. So the occurrence rate of WLFs in X-class CFs is 100\% in our sample. For the M-class CFs, the occurrence rate of WLFs is $\sim$62\%, which is still much higher than the occurrence rate of WLFs found in previous investigations. However, only 4 CFs with a C class are WLFs, corresponding to an occurrence rate of $\sim$8.5\%. The only B-class CF in our sample does not reveal any detectable WL enhancement. 

Some ARs have each produced several CFs and all of them are WLFs, such as NOAA ARs 11476, 11890 and 12297. Each of NOAA ARs 12268, 12434 and 12497 has also produced several CFs. However, none of them is a WLF. Why are some ARs cradles of WLFs and others not? We realize that CFs in the former group of ARs are generally much stronger than those in the latter group, suggesting that CFs in the former group of ARs are more energetic. In addition, the sizes of flaring regions surrounded by the circular flare ribbons of CFs in NOAA AR 12268 are about 3 or 4 times larger than those in the former group of ARs, implying that the spatial scale might be an important factor for the production of WLFs.

\subsection{Comparison of WL CFs with non-WL CFs }

\subsubsection{ Area surrounded by circular ribbon}

Figure 2 presents two examples of CFs. One is an M5.3 flare occurring on 2011 September 6th in NOAA AR 11283, which has a big circular ribbon and is a non-WL flare. The other event is an X3.3 flare occurring on 2013 November 5th in NOAA AR 11890, and the area enclosed by the circular ribbon is small. This flare is a WLF. The WL enhancement is relatively large, with an average value ($dI_{wl}^a$) of $\sim$0.22 and a maximum value ($dI_{wl}^m$) of $\sim$1.05. The red circles shown in the middle and right panels mark the outer edges of the circular ribbons in the 1600 \AA\ images. We can see that the HMI magnetograms are clearly characterized as one magnetic polarity surrounded by the opposite-polarity magnetic fluxes. For each CF we then calculated the area enclosed by the red circle, which is defined as the area of flaring region ($S_{cr}$). Deprojection has been done before the calculation.

Figure 4(a) shows the scatter plot of the relationship between the \textit{GOES} 1-8 \AA\ soft X-ray peak fluxes and the areas of flaring region ($S_{cr}$) for these 90 CFs. As mentioned previously, all the X-class CFs are WLFs, regardless of the different values of $S_{cr}$. For the M-class CFs, WLFs predominantly have smaller values of $S_{cr}$. The average value of $S_{cr}$ for the M-class WL CFs is about 1823 $arcsec^2$. As a comparison, the average value of M-class non-WL CFs is about 3865 $arcsec^2$. Obviously, M-class CFs with a smaller flaring region are more likely to be WLFs. For C- and B-class CFs, most of them do not reveal any obvious enhancement in the WL emission, even if the area of flaring region is small. Only 4 C-class CFs are WLFs, and the average area of their flaring regions is about 863 $arcsec^2$. An extreme case is the C2.3 WL CF in NOAA AR 12615 marked by the green arrow in Figure 4(a). The area of the flaring region in this CF is only 194 $arcsec^2$, which is the smallest in our sample. For non-WL C-class CFs, the average value of $S_{cr}$ is about 1435 $arcsec^2$. 

The fact that CFs showing WL enhancement tend to be associated with smaller circular ribbons may have important implications for the energy release in WLFs. In the fan-spine magnetic field topology, a small circular ribbon likely indicates a low-lying null point or quasi-separatrix layer. Under such a situation, magnetic reconnection and the subsequent energy release should occur at low heights. As a result, the released energy may penetrate to the even lower atmosphere more easily, and produce enhanced WL emission there. In addition, a smaller flaring region is equivalent to a larger energy flux if the released energy is the same. Obviously, a larger energy flux will more likely impact the lower atmosphere and produce excess WL emission. We noticed that Watanabe et al. (2017) found an association of WLFs with shorter ribbon distances, which may also indicate lower reconnection heights and thus is consistent with our finding. 

\subsubsection{ Flare duration}

We have calculated two durations for each CF. One is the flare duration ($\Delta T$). The beginning time is defined as the time when the \textit{GOES} 1-8 \AA\ soft X-ray flux starts to show a steep monotonic increase. The ending time refers to the time when the \textit{GOES} flux decays to a level halfway between the maximum and the pre-flare background level. Another duration is the duration of the impulsive phase ($\Delta T^\prime$). We first took the time derivative of the \textit{GOES} 1-8 \AA\ flux, then defined $\Delta T^\prime$ as the period when the derivative is above 1/$e$ of the maximum value. Figure 3 demonstrates how these two durations were calculated. The impulsive phase duration ($\Delta T^\prime$) basically reflects how fast the energy is released in the impulsive phase. Comparing to the whole flare duration ($\Delta T$), the impulsive phase duration ($\Delta T^\prime$) is usually much shorter.

The values of $\Delta T$ and $\Delta T^\prime$ for all the 90 CFs are plotted in Figure 4(b) and (c). There is a clear trend that WLFs tend to have shorter durations. This is especially true for the M-class flares. The average values of $\Delta T$ and $\Delta T^\prime$ for the M-class WL CFs are 808 s and 195 s, respectively. As a comparison, these two values are 1426 s and 244 s for M-class non-WL CFs, respectively. For the small flares (C-class and B-class), only four of them are WLFs and their durations are much smaller than the average durations. An extreme case is the C2.3 flare occurring in NOAA AR 12615. This WLF is marked by a green arrow in Figure 4(b) and (c). Its durations are 300 s ($\Delta T$) and 18 s ($\Delta T^\prime$), the shortest among all the 90 CFs. 

In the statistical investigation of M- and X-class flares observed with \textit{Hinode}/SOT, Watanabe et al. (2017) found that WLFs are generally characterized by a shorter timescale. They concluded that the precipitation of many nonthermal electrons during a short time period likely play a crucial role in the production of WL enhancement. Here through a detailed investigation of CFs, we find a similar result and thus a similar conclusion may be drawn. Given the same amount of released energy, a shorter time scale means a larger rate of energy precipitation. Under such a situation, the deposited energy may lead to rapid heating of the lower atmosphere and produce enhanced WL emission more easily.

\subsubsection{ Magnetic field and electric current in the region surrounded by circular ribbon}

We also examined the possible difference of magnetic field and electric current between WLFs and non-WL flares. The calculation of the electric current density perpendicular to the photosphere ($J_z$) is based on the Ampere$^\prime$s law:

\begin{equation}
\emph{$J_{z}=\displaystyle\frac{1}{\mu_0}(\triangledown \times \textbf{B})_{z}=\frac{1}{\mu_0}(\frac{\partial B_y}{\partial x}-\frac{\partial B_x}{\partial y})$}\label{equation1},
\end{equation}
where $\mu_0$ is the magnetic permeability of vacuum, $B_x$ and $B_y$ are the two horizontal components of the photosphere vector magnetic field measured by HMI. It should be noted that we used the hmi.sharp\_cea\_720s vector magnetic field, which has been processed by a cylindrical equal area (CEA) projection. The coordinates (x, y, z) here refer to the heliographic longitude, latitude and radial directions. 

Figure 5 shows spatial distribution of the vector magnetic field and radial electric current ($J_z$) in AR 11476 before an M5.7 WLF and in AR 12434 before an M1.1 non-WL flare. The vector magnetic field in the flaring region of AR 11476 is very complicated. The horizontal magnetic field in the main negative-polarity sunspot possesses a rotating pattern, which implies a strong twist of the magnetic field. The corresponding electric current density is very strong. The WL enhancements are generally located in regions with a large current density (Figure 5(b)). In the flaring region of AR 12434, the vector magnetic field is less complicated and the electric current density is much weaker than that of AR 11476. It should be noted that AR 11476 has produced 5 CFs, all of which are WLFs. Though AR 12434 has produced 8 CFs, none of them is a WLF. This difference might be related to the difference of magnetic field complexity and electric current density in the two ARs.

The radial components of the unsigned magnetic field strength ($|B_z|$) and electric current ($|J_z|$) integrated over the region enclosed by each circular ribbon are presented in Figure 7(a) and (b), respectively. If we only consider CFs with a \textit{GOES} class lower than X, we find no big difference of $|B_z|$ between WLFs and non-WL flares, though a weak trend of stronger $|B_z|$ in WLFs appears to be present. However, if the average magnetic field is very weak, i.e., $|B_z|<400G$, then CFs are very unlikely to be WLFs. The electric current $|J_z|$ appears to be distinctly different between WLFs and non-WL flares. We see a clear trend that WLFs are associated with larger $|J_z|$ values, suggesting that the nonpotentiality plays an important role in the generation of WL enhancement. 

\subsubsection{ Hard X-ray power-law index}

The hard X-ray (HXR) power-law index could reflect the process of electron acceleration during a flare. A larger power-law index (absolute value) usually means that the spectrum is relatively soft with less high-energy electrons. While a smaller power-law index means that the spectrum is relatively hard and that there are a larger number of high-energy electrons. In our sample, 50 CFs were observed simultaneously by \textit{RHESSI}. Among these 50 flares, 21 are WLFs (4 X-class, 16 M-class and 1 C-class flares) and 29  are non-WL flares (9 M-class, 19 C-class and 1 B-class flares). Following Huang et al. (2016), we fitted the \textit{RHESSI} spectra at or around the peak times of the flares by using the model of a variable thermal function (vth) plus a broken power law function (bpow). As an example, Figure 6 shows the fitting to the \textit{RHESSI} spectrum obtained during an M4.7 flare on 2012 May 9th in NOAA AR 11476.

Figure 7(c) shows the HXR power-law indexes and the peak values of GOES flux for these 50 CFs. Among these 50 CFs, all the X-class CFs are WLFs, and almost all the C- and B-class CFs are not WLFs. For the M-class CFs, we see no obvious difference of the HXR power-law index between WLFs and non-WL flares. A similar result has also been found by Watanabe et al. (2017), though the flares they analyzed are not CFs. 

\subsubsection{ Association with CMEs}

Based on the association with CMEs, flares are classified as ``eruptive flares" and ``confined flares" (e.g., \v{S}vestka \& Cliver 1992). By checking the \textit{Solar and Heliospheric Observatory (SOHO)}/LASCO CME catalog (Yashiro et al. 2004; https://cdaw.gsfc.nasa.gov/CME\_list/) and examining UV and EUV observations by AIA using the JHelioviewer software (M\"uller et al. 2017), we find 61 confined flares and 29 eruptive flares in our sample. And among the 33 WLFs, 18 are confined flares and 15 are eruptive flares. Thus, it appears that WLFs do not have a preference to be eruptive flares or confined flares. 

\subsection{WL enhancement in the CFs}

More than half of the 90 CFs are accompanied by filament eruptions. Among the 33 WLFs, 20 are obviously associated with filament eruptions (11 of them have accompanied CMEs and the other 9 have no accompanied CMEs), whereas the other 13 are not. For the WLFs accompanied by filament eruptions, the WL enhancement could appear at the footpoints of the erupting filaments (Figure 8(b) and (f)), below or in the vicinity of the filaments (Figure 8(b)). These WL enhancements are likely related to the eruption of filaments. No matter there is a filament eruption or not, all the WL enhancements in our sample appear inside the circular flare ribbons (Figure 8(c), (g) and (k)) or on the flare ribbons (Figure 8(c) and (g)).

Figure 9 shows the relationship between the WL enhancement and the peak value of \textit{GOES} soft X-ray flux. For each WLF, we define a WLF region as the area where the WL enhancement ( $(I_{p}-I_{0})/I_{0}$) is greater than 0.05. Here $I_p$ and $I_0$ represent the HMI continuum intensities at the two times when the WL emission reaches the peak and before occurrence, respectively. The average ($dI_{wl}^a$) and maximum ($dI_{wl}^m$) values of WL enhancement in the WLF regions are presented on the left and right panels of Figure 9, respectively. For the X-class WL CFs, there appears to be a positive correlation between the WL enhancement and the peak of \textit{GOES} soft X-ray flux. Higher-class CFs tend to have stronger WL enhancement, indicating that the WL enhancement is largely related to the total amount of released energy. However, for the M- and C-class WL CFs, we see no obvious correlation between the magnitude of WL enhancement and the flare class. The absence of correlation possibly suggests that the energy released during these relatively small flares is not a dominated factor to determine the WL enhancement. Other factors such as the time scale, spatial scale and the height of energy release may also play equally important roles in the generation of WL enhancement.

Figure 10 shows the relationship between WL enhancement and other parameters including the area of flare region ($S_{cr}$), HXR power-law index, flare duration ($\Delta T$ and $\Delta T^\prime$), radial component of the magnetic field strength ($|B_z|$) and electric current ($|J_z|$). There seems to be no strong correlation between the WL enhancement and these parameters, though there is a weak tendency that a greater WL enhancement more likely occurs in flares with shorter impulsive phase durations (Figure 10(d)). Possibly, the magnitude of WL enhancement is determined by a combination of different parameters rather than a single parameter. It is also possible that the main parameter that determines the magnitude of WL enhancement is different for different flares.


\section{Summary}

We have performed the first statistical study of CFs and investigated the WL emission in these flares. Our aim is to examine whether CFs, which are usually associated with a fan-spine magnetic field topology, are more likely to show WL enhancement. We have also studied the differences between WL CFs and non-WL CFs. From nearly 8-year observations of \textit{SDO}, we have identified 90 CFs from 36 ARs, including 8 X-class flares, 34 M-class flares, 47 C-class flares and 1 B-class flare. Among these 90 CFs, 33 of them are WLFs, including 8 X-class CFs, 21 M-class CFs and 4 C-class CFs. Thus, the occurrence rate of WLFs is about 37\% (33/90) for CFs. The occurrence rate is even larger, about 69\% (29/42), for CFs greater than M1.0. This is much higher than previously reported occurrence rates, suggesting that the fan-spine magnetic field topology favors the occurrence of WLFs. However, only about 8\% (4/48) of the C- and B-class CFs are WLFs. It is also worth noting that some ARs have each produced several CFs and nearly all of them are WLFs. 

We have derived some parameters, including the area of the region surrounded by the circular ribbon ($S_{cr}$), flare duration ($\Delta T$ and $\Delta T^\prime$), radial component of the photospheric magnetic field strength ($|B_z|$) and electric current ($|J_z|$), and HXR power-law index. We have investigated the differences of these parameters between WL CFs and non-WL CFs. Our analysis suggests that the CFs with WL enhancement tend to have a smaller spatial scale, shorter duration, stronger and more complicated magnetic field. 

We find that for X-class WL CFs, the WL enhancement has a positive correlation with the flare class. This result, together with the fact that all the identified X-class CFs are WLFs, suggests that the magnitude of WL enhancement in large flares is largely determined by the amount of released energy. However, for M- and C-class WL CFs, there is no obvious correlation between the WL enhancement and flare class. This suggests that other factors such as the time scale, spatial scale and magnetic field complexity may play important roles in the generation of WL emission if the released energy is not high enough. For instance, if the released energy is low (C- and B-class), and the spatial and temporal scales are extremely small, the released energy may still produce rapid and efficient heating of the lower atmosphere, leading to detectable WL enhancement. 

We noticed that in other magnetic field configurations even some X-class flares may not be WLFs (e.g., Watanabe et al. 2017). This observational fact, together with our findings about CFs, suggest that the energy as measured by GOES classification is not a sufficient condition for the production of WL emission. A right magnetic field configuration may  also be needed as well as enough energy. The GOES energy is likely a secondary consideration for the production of WLFs of all classes. Of course, it should be noted that the energy as measured by GOES is only a part of the flare energy budget.

\acknowledgements
This work is supported by the Strategic Priority Research Program of CAS with grant XDA17040507, CAS Key Laboratory of Solar Activity (No:KLSA201810, National Astronomical Observatories of CAS), the Max Planck Partner Group program and the Recruitment Program of Global Experts of China. We thank Prof. Minde Ding and Prof. Jun Zhang for helpful discussion.

\newpage

\begin{table*}[htbp]
\centering
\begin{threeparttable}
\centerline{\footnotesize Table 1. List of circular-ribbon flares}
\label{tab1}
{\tiny
\begin{tabular}{lccccccrrrccccc}
\hline
\hline
NOAA &Date&Peak&GOES& Sunspots &Flare &$\Delta T$&$\Delta T^\prime$&$S_{cr}$&$|B_z|$&$|J_z|$&HXR &$dI_{wl}^a$ &$dI_{wl}^m$&CME\\
AR&&Time&Class& Location&Ribbon&(min)&(min)&(${1^{\prime\prime}}^2$)&(G)&($mA/m^2$)&Index& & \\
\hline
11283&	                 2011.09.06&	01:50&	M5.3	&      N14W07&	C&	 28.0&   3.93&	5425.13&  366.23&	19.47&	6.49&	...&	...&	Yes\\
$11283\bigstar$&	2011.09.06&	22:20&	X2.1&	N14W18&	C&	 12.0&   1.70&   4958.17&  456.58&	21.68&	3.10&	0.169 &0.793&	Yes\\
$11283\bigstar$&	2011.09.07&	22:38&	X1.8&	N14W28&	C&	 12.0&   1.92&  3860.86&  556.76&	24.29&	...&	0.119	&0.449&	Yes\\
$11283\bigstar$&	2011.09.08&	15:46&	M6.7&	N14W40&	C&	 20.0&   4.53&	4318.15&  438.94&	16.60&	3.08&	0.062	&0.081&	Yes\\							
11324&	                 2011.10.22&	15:20&	C4.1&	N11E18& C&	 15.0&   3.25&   2658.74&  409.71&	32.52&	10.00&	...&...&	No\\
$11339\bigstar$&	2011.11.03&	20:27&	X1.9&	N22E63&	C&    16.0&    5.11&   1313.29&  1121.95&22.25&	...&	0.106&0.520&	Yes\\
11339&	                 2011.11.06&	06:35&	M1.4&	N21E31& C&    27.0&    1.88&  929.30&  729.23&	28.50&	...&	...&...&	No\\
11339&	                 2011.11.06&	09:56&	C8.8&	N21E28&	C&	   7.0&    1.53&  1140.90&  464.50&	18.57&	...&	...&...&	No\\
11339&	                 2011.11.06&	14:47&	C5.3&	N21E27&	QC&  33.0&    2.02&  768.29&  838.00&	37.41&	5.50&	....&...&	No\\
$11346\bigstar$&	2011.11.15&	12:43&	M1.9&	S17E30&	QC&	  20.0&   3.03&  2452.57&  600.19&	26.33&	3.71&	0.056&	0.059&	Yes\\
11346&	                 2011.11.16&	13:37&	C2.8&	S18E19&	QC&  16.0&    2.70&  777.91&  291.62&	18.91&	11.50&	...&...&	Yes～\\
11346&	                 2011.11.16&	18:54&	C2.9&	S18E16&	QC&  11.0&    1.95&   662.45&  207.87&	18.69&	...&	...&...&	Yes～\\
11346&	                 2011.11.16&	23:48&	C5.0&	S12E20&	QC&	 17.0&    3.78&   2675.68&  370.29&	  20.52&	3.28&	...&...&	Yes～\\
$11476\bigstar$&	2012.05.08&	13:08&	M1.4&	N13E44&	C&	 10.0&     2.02&  1051.28&  1162.24&  38.07&	4.88&	0.081&0.134&	No\\
$11476\bigstar$&	2012.05.09&	12:32&	M4.7&	N13E31&	C&	 15.0& 	6.00& 1323.93&  1124.55&  41.93&	3.86&	0.081&	0.156&	No\\
$11476\bigstar$&	2012.05.09&	21:05&	M4.1&	N12E26& C&	   8.0&	1.95&  1236.27&  1070.47&  44.49&	...&	0.074&0.119&	No\\
$11476\bigstar$&	2012.05.10&	04:18&	M5.7&	N12E22& C&	 12.0&	1.70&  1161.81&  959.21&	   40.35&	3.97&	0.089&0.222&	No\\
$11476\bigstar$&	2012.05.10&	20:26&	M1.7&    	N12E12&	C& 	10.0&	3.38&  1276.14&  478.12&   22.51&	4.76&	0.071&0.096&	No\\
$11598\bigstar$&	2012.10.22&	18:51&	M5.0&	S12E61&  C&   	23.0&	2.08&  2192.78&  ...&	      ...&	...&	0.081&0.144&	No\\
$11598\bigstar$&	2012.10.23&	03:17&	X1.8&	S13E58&	C&	  8.0&	1.37&  1637.57&  1831.46&  38.35&	3.64&	0.231&1.153&	No\\
$11652\bigstar$&	2013.01.13&	08:38&	M1.7&	N17W22&	C&      5.0&       1.02&  704.92&  536.20&	36.00&	...&	0.066&0.100&	Yes～\\
11652&	                 2013.01.14&	01:22&	C6.5&	N18W31&	QC&	 10.0&	4.07&  653.14&  557.48&	 42.86&	...&	...&...&	Yes～\\
11669&	                 2013.02.05&	06:12&	B6.6&	N08E63& QC&    8.0&      1.92&  2282.97&  388.99& 12.38&	9.41&	...&...&	No～\\
11669&	                 2013.02.05&	08:19&	C6.3&	N07E64&	C&      10.0&     4.75&  2213.47&  397.15&  11.16&	...&	...&...&	Yes～\\
$11675\bigstar$&	2013.02.17&	15:50&	M1.9&	N12E22&	C&       7.0&       0.78&  136.16&  1553.62& 79.35&	7.39&	0.169&0.397&	No\\
11689&	                 2013.03.12&	22:46&	C3.6&	S21W41&	C&	    7.0&       1.20&  158.23&  476.38&	     28.53&	...&	...&...&	No\\
11731&	                 2013.04.28&	15:55&	C3.7&	N10E29&	C&	    6.0&	 3.97&  576.63&  969.64&	 40.66&	...&	...&...&	No\\
11731&	                 2013.04.28&	17:02&	C1.8&	N10E28&	C&       6.0&       1.53& 700.24&  629.81&	30.00&	...&	...&...&	No\\
11731&	                 2013.05.02&	05:10&	M1.1&	N10W26&	C&      21.0&      5.87& 7516.71&  169.38&   8.94&	1.65&	...&...&	Yes～\\
$11890\bigstar$&	2013.11.05&	08:18&	M2.5&	S16E51&	C&	   9.0	&       2.50& 1343.94&  807.75&   30.06&	...&	0.072&0.110&	No\\
$11890\bigstar$&	2013.11.05&	22:12&	X3.3&	S12E44&	C&      8.0&        1.43& 1435.23&  1276.58&	34.98&	...&	0.220&1.050&	Yes\\
$11890\bigstar$&	2013.11.06&	08:51&	C8.6&	S13E38&	C&     10.0&        3.17& 1334.40&  1778.48&	48.87&	4.91&	0.076&0.145&	No\\
$11890\bigstar$&	2013.11.06&	13:46&	M3.8&	S12E35&	C&     14.0&        2.22&  1468.64&  1268.70&	45.21&	3.35&	0.100&0.270&	Yes～\\
$11890\bigstar$&	2013.11.07&	03:40&	M2.3&	S13E28&	C&     9.0&          2.53&  1348.55&  1035.92&	36.50&	3.78&	0.077&0.156&	No\\
$11890\bigstar$&	2013.11.07&	12:29&	C5.9&	S13E23&	C&     12.0&        2.32&  1538.22&  838.90&	33.21&	...&	0.065&0.084&	No\\
$11890\bigstar$&	2013.11.08&	04:26&	X1.1&	S13E13&	C&    9.0&           1.43&  2522.65&  486.71&	22.30&	...&	0.125&0.622&	Yes\\
$11890\bigstar$&	2013.11.10&	05:14&	X1.1&	S13W13&	C&    10.0&         1.77&  4238.09&  423.40&	17.21&	4.44&	0.116&0.259&	Yes\\
11936&	                 2013.12.28&	12:47&	C3.0&	S17E09&	C&	18.0	&         2.87&  1358.40&  386.70&	17.07&	...&	...&...&	No\\
11936&	                 2013.12.28&	18:02&	C9.3&	S17E07&	C&	15.0	&        4.57&  1229.78&  442.91&	19.77&	6.95&	...&...&	No\\
$11936\bigstar$&	2013.12.29&	07:56&	M3.1&	S16W01&	C&	11.0	&        2.70&  1405.61&  567.34&	26.22&	...&	0.066&0.101&	No\\
11936&	                 2013.12.29&	14:46&	C5.1&	S16W05&	C&	11.0	&        2.90&  1016.23&  405.92&	18.35&	2.20&	...&...&	No\\
$11936\bigstar$&	2014.01.01&	18:52&	M9.9&	S16W45&	QC&	 23.0	&        4.23&  6973.87&  460.18&	13.20&	2.36&	0.066&0.117&	Yes\\
11991&	                 2014.03.05&	00:16&	C4.8&	S27W07&	C&     9.0&          2.28&  348.49&  1278.55&	69.95&	2.48&	...&...&	No\\
11991&	                 2014.03.05&	01:58&	C2.8&	S27W08&	C&	  9.0	&        3.68&  402.24&  1025.70&	63.00&	2.67&	...&...&	No\\
11991&	                 2014.03.05&	02:10&	M1.0&	S27W08&	C&	  6.0	&        3.17&  407.57&  1158.73&	65.73&	2.04&	...&...&	Yes～\\
12017&	                 2014.03.28&	19:18&	M2.0&	N10W20&	QC&	  23.0&       4.57&  3586.71&  371.80&	17.67&	3.55&	...&...&	Yes～\\
$12017\bigstar$&	2014.03.28&	23:51&	M2.6&	N10W22&	C&	  14.0 &      3.45&  3081.18&  754.17&	37.14&	2.51&	0.070&0.125&	Yes\\
$12017\bigstar$&	2014.03.29&	17:48&	X1.0&	N10W32&	C&	  18.0&      1.85&  5281.42&  386.29&	18.61&	2.76&	0.084&0.173&	Yes\\
12031&	                 2014.04.06&	21:01&	C3.8&	N02W23&	QC&   17.0&      3.07&  2530.27&  470.80&	23.50&	8.77&	...&...&	No\\
12035&	                 2014.04.15&	09:23&	C8.6&	S14E25&C&      10.0&      2.70&  1403.33&  260.34&	16.51&	...&	...&...&	Yes～\\
12035&	                 2014.04.15&	17:59&	C7.3&	S15E21&C&      10.0&      1.73&  3365.81&  134.22&	10.86&	...&	...&...&	Yes～\\
12036&	                 2014.04.18&	13:03&	M7.3&	S20W34&	QC&    49.0&    11.33&  41509.00&  228.28&	7.45&	2.52&...&...&	Yes	\\								
														
\hline
\end{tabular}
}
\end{threeparttable}
\end{table*}

\begin{table*}[htbp]
\centering
\begin{threeparttable}
\centerline{\footnotesize Table 1. (Continued)}
\label{tab1}
{\tiny
\begin{tabular}{lccccccrrrccccc}
\hline
\hline
NOAA &Date&Peak&GOES& Sunspots &Flare&$\Delta T$&$\Delta T^\prime$&$S_{cr}$&$|B_z|$&$|J_z|$&HXR&$dI_{wl}^a$&$dI_{wl}^m$&CME\\
AR&&Time&Class& Location &Ribbon&(min)&(min)&(${1^{\prime\prime}}^2$)&(G)&($mA/m^2$)&Index& & \\
\hline
12087&	                 2014.06.13&	07:56&	M2.6&	S18E40&QC&	   10.0&     2.38&  1293.73&  897.95&	19.65&	7.56&	...&...&	No\\
12087&	                 2014.06.13&	20:17&	C9.0&	S20E34&	QC&	   7.0	&       1.98&  1128.34&  1105.80&	27.77&	...&	...&...&	No	\\							
12148&	                 2014.08.22&	00:06&	C6.6&	N07E31&	C&	   16.0&     1.47&  838.08&  690.23&	     32.70&	         3.73&	...&...&	No		\\									
12146&	                 2014.08.22&	10:27&	C2.2&	N12E01&	C&	   33.0&    10.58&  2159.82&  228.29&	12.13&	...&	...&...&	No		\\									
12157&	                 2014.09.13&	12:54&	C3.7&	S16W38&	C&	   6.0	&       2.12&  389.49&  ...&	                     ...&	1.96&	...&...&	No		\\							
$12192\bigstar$&	2014.10.20&	19:02&	M1.4&	S15E46&	C&	   9.0	&       3.78&  729.09&  569.31&	      27.52&	3.41&	0.064&0.087&	No～	\\									
12201&	                 2014.11.03&	03:52&	C4.2&	S03E21&	QC&	   9.0	&       2.32&  1421.64&  249.04&	19.25&	...&	...&...&	No～	\\								
12227&	                 2014.12.13&	10:10&	C4.0&	S06W67&	C&	  13.0&       3.97&  1821.88&  854.73&	24.18&	...&	...&...&	No～		\\				
12266&	                 2015.01.19&	20:48&	C3.3&	S07E06&	C&	  11.0&        4.03&  751.06&  103.46&	6.91&	...&	...&...&	No	\\										
12268&	                 2015.01.29&	05:23&	C8.2&	S13W06&	QC&	  35.0&        4.92&  6680.02&  772.63&	27.28&	2.65&	...&...&	No\\
12268&	                 2015.01.29&	11:42&	M2.1&	S12W06&	QC&	  20.0&        2.90&  6509.92&  785.76&	25.52&	5.93&	...&...&	No\\
12268&	                 2015.01.29&	19:53&	C6.4&	S13W13&	QC&	  53.0&        2.15&  5605.85&  835.53&	31.51&	2.33&	...&...&	No\\
12268&	                 2015.01.30&	00:44&	M2.0&	S13W16&	QC&  30.0&         6.35&  8556.89&  671.08&	26.58&	...&	...&...&	No\\
12268&	                 2015.01.30&	05:36&	M1.7&	S12W19&	QC&  66.0&         2.32&  9112.60&  766.64&	27.81&	...&	...&...&	No	\\						
12276&	                 2015.01.30&	08:25&	C3.8&	S06E08&	C&	 10.0	&         3.35&  1589.00&  216.75&	13.67&	...&	...&...&	Yes～		\\							
12277&	                 2015.02.03&	10:53&	C3.9&	N06W03&	C&	 14.0	&        4.47&  1030.82&  478.15&	30.71&	...&	...&...&	No～		\\			
$12297\bigstar$&	2015.03.10&	03:24&	M5.1&	S15E40&	C&     9.0&          2.12&  2024.09&  798.95&	34.71&	3.48&	0.101&0.227&	Yes\\
$12297\bigstar$&	2015.03.11&	00:02&	M2.9&	S16E28&	C&     20.0&        1.30&  1289.82&  712.00&	31.50&	5.41&	0.076&0.142&	Yes～\\
$12297\bigstar$&	2015.03.12&	21:51&	M2.7&	S15E01&	C&    12.0&         1.82&  1499.89&  800.38&	41.14&	2.03&	0.058&0.080&	No\\
$12297\bigstar$&	2015.03.13&	06:07&	M1.8&	S14W02&	C&    23.0&       15.02&  1274.46&  487.08&	28.44&	12.40&	0.062&0.108&	No～	\\			
12325&	                 2015.04.16&	19:18&	C3.3&	N05E51&	C&    27.0&         5.50&  2056.98&  457.21&	17.71&	...&	...&...&	No			\\								
12434&	                 2015.10.15&	18:38&	C3.9&	S11E52&	C&    10.0&         3.75&  1131.89&  1493.97&	40.12&	...&	...&...&	No\\
12434&	                 2015.10.15&	19:56&	C3.4&	S11E54&	C&    8.0&           3.13&  882.86&  1143.51&	31.29&	...&	...&...&	No\\
12434&	                 2015.10.15&	21:41&	C3.1&	S11E52&	C&   14.0&          1.82&  922.24&  795.32&	12.71&	4.10&	...&...&	No\\
12434&	                 2015.10.15&	23:31&	M1.1&	S11E50&	C&   10.0&          3.42&  1169.83&  1130.73&	28.41&	6.99&	...&...&	No\\
12434&	                 2015.10.16&	06:16&	M1.1&	S11E46&	C&   9.0&            2.12&  1191.66&  933.22&	24.47&	4.12&	...&...&	No\\
12434&	                 2015.10.16&	09:03&	C3.4&	S12E45&	C&   9.0&            2.45&  1199.57&  1039.83&	31.14&	4.02&	...&...&	No～\\
12434&	                 2015.10.16&	10:20&	C3.1&	S11E43&	C&   11.0&          3.00&  1205.01&  940.79&	28.12&	3.11&	...&...&	No～\\
12434&	                 2015.10.16&	13:42&	C4.2&	S11E41&	C&   10.0&         3.42&  1249.23&  913.74&	26.59&	3.60&	...&...&	Yes～	\\									
12497&	                 2016.02.13&	08:31&	C1.3&	N13W25&	C&   9.0&           4.10&  464.35&  627.71&	        39.45&	4.30&	...&...&	No～\\
12497&	                 2016.02.13&	10:09&	C2.8&	N13W25&	C&   9.0&            2.08&  512.41&  699.69&	41.91&	4.95&	...&...&	No\\
12497&	                 2016.02.13&	15:24&	M1.8&	N14W28&	C&	10.0	&        2.53&  681.78&  1089.25&	51.32&	...&	...&...&	No\\
12497&	                 2016.02.13&	18:55&	C1.6&	N14W30&	QC& 12.0&         0.58&  754.09&  1464.60&	70.63&	...& ...&...&	No\\
12497&	                 2016.02.14&	04:39&	C3.4&	N13W36&	QC&  25.0&        3.00&  891.52&  608.31&	30.81&	...&	...&...&	No～	\\								
$12567\bigstar$&	2016.07.16&	07:04&	C6.8&	N04E25&	C&    10.0&         1.63&  385.67&  403.30&	24.94&	...&	0.084&0.134&	No		\\						
$12615\bigstar$&	2016.11.30&	15:25&	C2.3&	S07E42&	QC&  5.0	&         0.30&  193.87&  890.13&	50.83&	...&	0.071&0.098&	No\\
12661&	2017.06.03&	09:57&	C2.1&	N06E55&	C&	                  28.0	&         2.70&  1147.90&  748.36&	18.88&	...&	...&...&	No～\\
12661&	2017.06.03&	14:56&	C2.5&	N06E52&	C&	                 13.0	&          2.18&  1209.89&  645.88&	20.86&	...&	...&...&	No～\\
\hline
\end{tabular}
}
\begin{tablenotes}
        \tiny
        \item[1] Flares marked with `$\bigstar$' refer to WLFs.
        \item[2] $\Delta T$ refers to the flare duration obtained from the \textit{GOES} 1-8 \AA\ X-ray flux.  $\Delta T^\prime$ refers to the impulsive phase duration of the flare based on the time derivative of the GOES flux. Please refer to the definitions of $\Delta T$ and $\Delta T^\prime$ in the text. $S_{cr}$ is the area of the region surrounded by the circular flare ribbon. $|B_z|$ and $|J_z|$ are the mean unsigned values of the radial component of the photospheric magnetic field and electric current, respectively. 
        \item[3]$dI_{wl}^{m}$ and $dI_{wl}^{a}$ are the WL enhancements calculated by $(I_{p}-I_{0})/I_{0}$. Here $I_{p}$ and $I_{0}$ are the HMI continuum intensities at two times when the WL emission reaches the peak and before appearance, respectively. $dI_{wl}^{m}$ and $dI_{wl}^{a}$ refer to the maximum and the average values of WL enhancement in a WLF region, respectively.                                                                                                                                                                                                              
        \item[4] For the flare ribbon, `C' means circular and `QC' means quasi-circular.
      \end{tablenotes}
\end{threeparttable}
\end{table*}

\clearpage

\newpage

\begin{figure}[!ht]
\centerline{\includegraphics[width=1\textwidth]{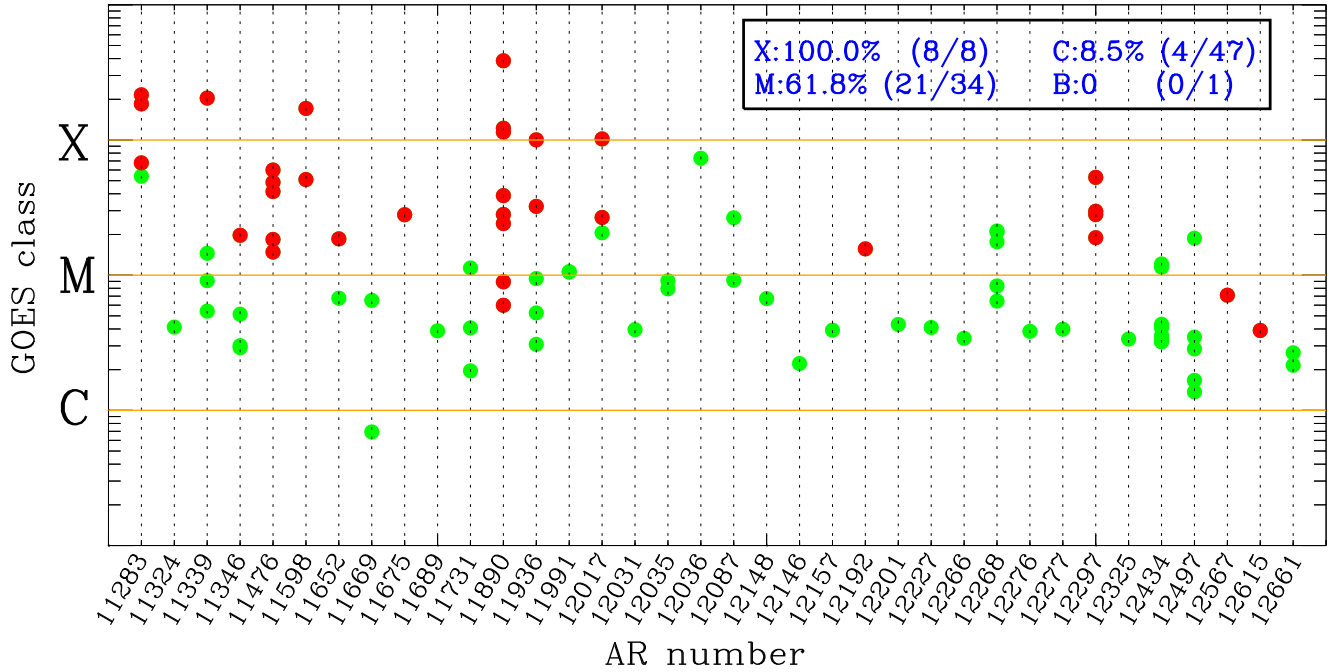}}
\caption{The hosting AR numbers and GOES classes of the 90 CFs. The red dots are WLFs, and the green ones are non-WL flares. The occurrence rates of WLFs with different GOES classes are printed in the black box.}
\end{figure}

\begin{figure}[!ht]
\centerline{\includegraphics[width=0.9\textwidth]{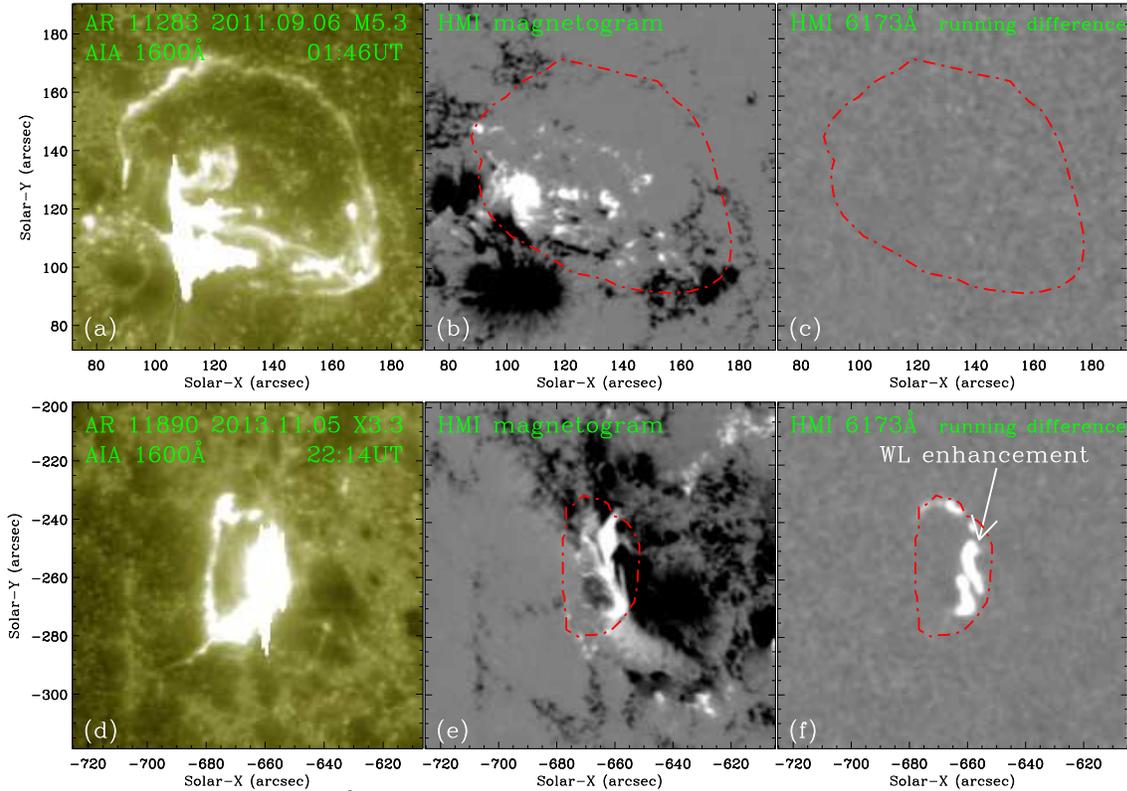}}
\caption{ The AIA 1600 \AA\ images, HMI line-of-sight magnetograms and HMI continuum difference images taken during two CFs. The red circles mark the outer edges of the circular ribbons in the 1600 \AA\ images. Upper: a non-WL flare, and the difference image (c) is taken betwen the peak time and the start time of the flare. Bottom: a WLF, and the difference image (f) is taken between the peak time and the start time of WL enhancement.}
\end{figure}

\begin{figure}[!ht]
\centerline{\includegraphics[width=0.9\textwidth]{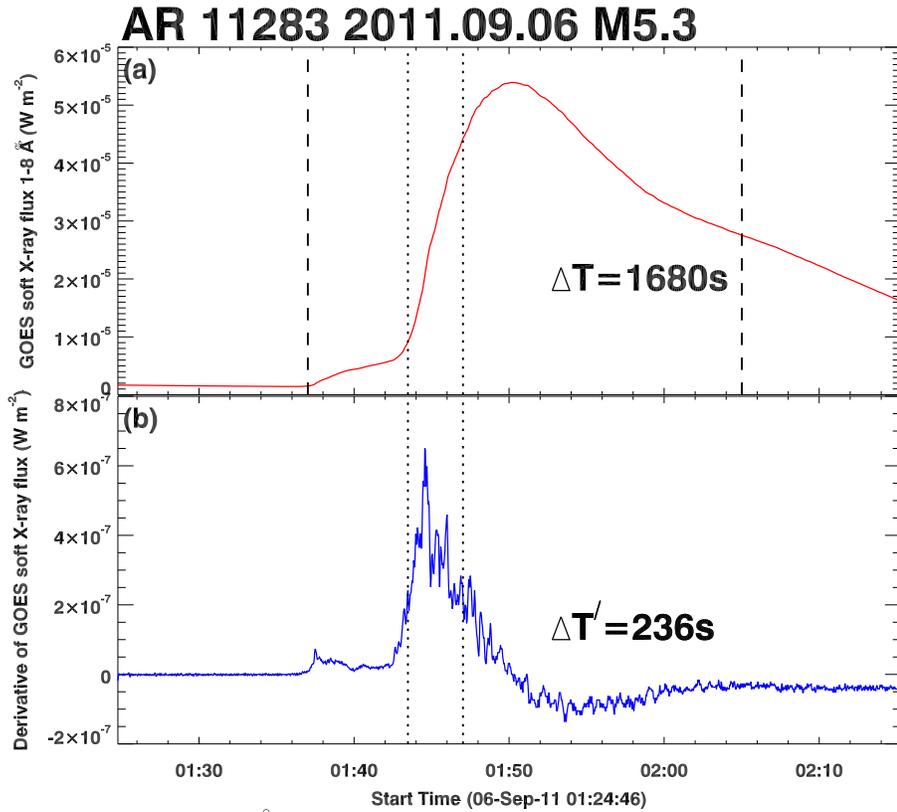}}
\caption{ The \textit{GOES} 1-8 \AA\ X-ray flux and its time derivative for an M5.3 flare occurring on 2012 September 6th in NOAA AR 11283. The two dashed lines in panel (a) define the duration of the flare ($\Delta T$). The two dotted lines define the impulsive phase duration ($\Delta T^\prime$). } 
\end{figure}

\begin{figure}[!ht]
\centerline{\includegraphics[width=0.9\textwidth]{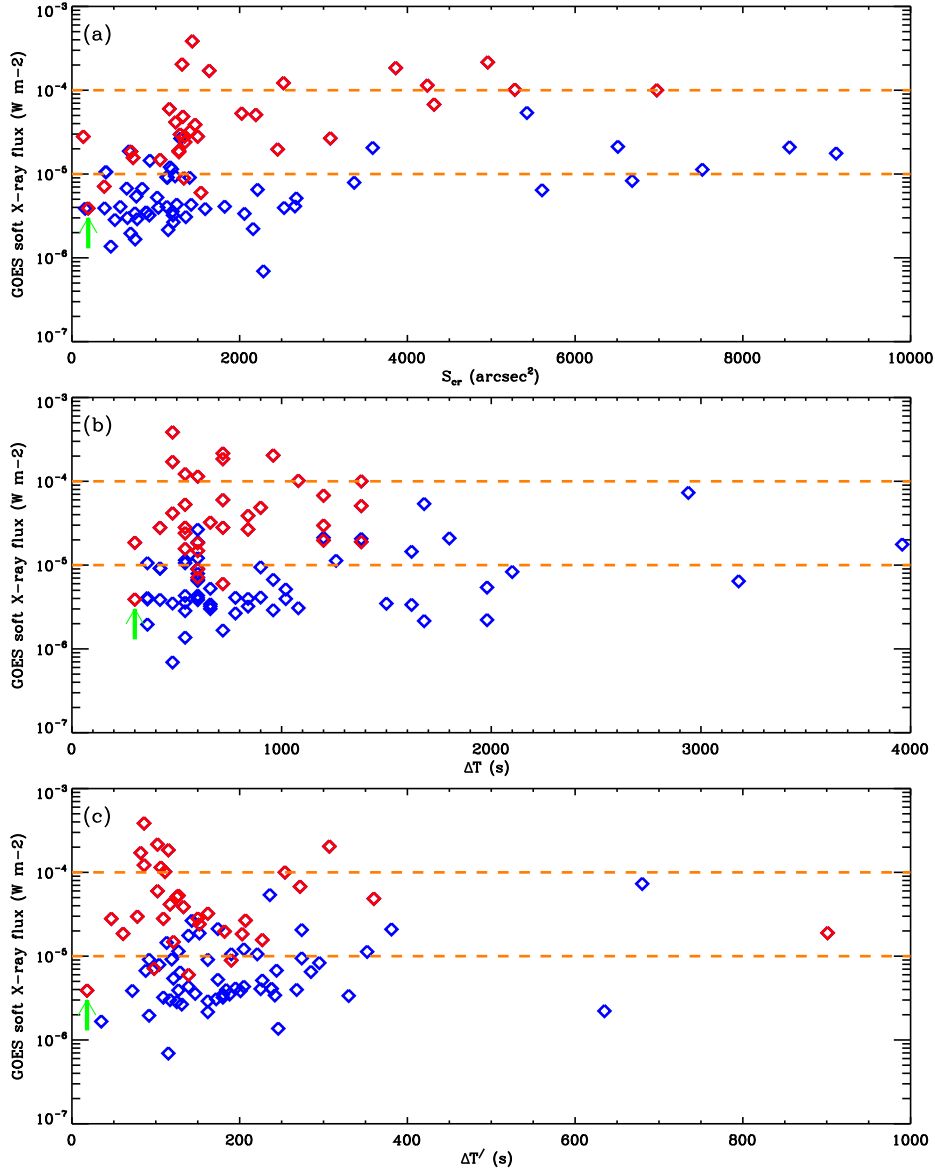}}
\caption{ Scatter plots showing the relationship between the peak of \textit{GOES} 1-8 \AA\ X-ray flux and the area enclosed by the circular ribbon ($S_{cr}$, a), the flare duration ($\Delta T$, b), and the impulsive phase duration ($\Delta T^\prime$, c). The red and blue diamonds represent WLFs and non-WL flares, respectively. The green arrows mark a C2.3 WLF occurring in NOAA AR 12615. The two brown horizontal lines indicate X1.0 and M1.0 classes, respectively.}
\end{figure}

\begin{figure}[!ht]
\centerline{\includegraphics[width=0.9\textwidth]{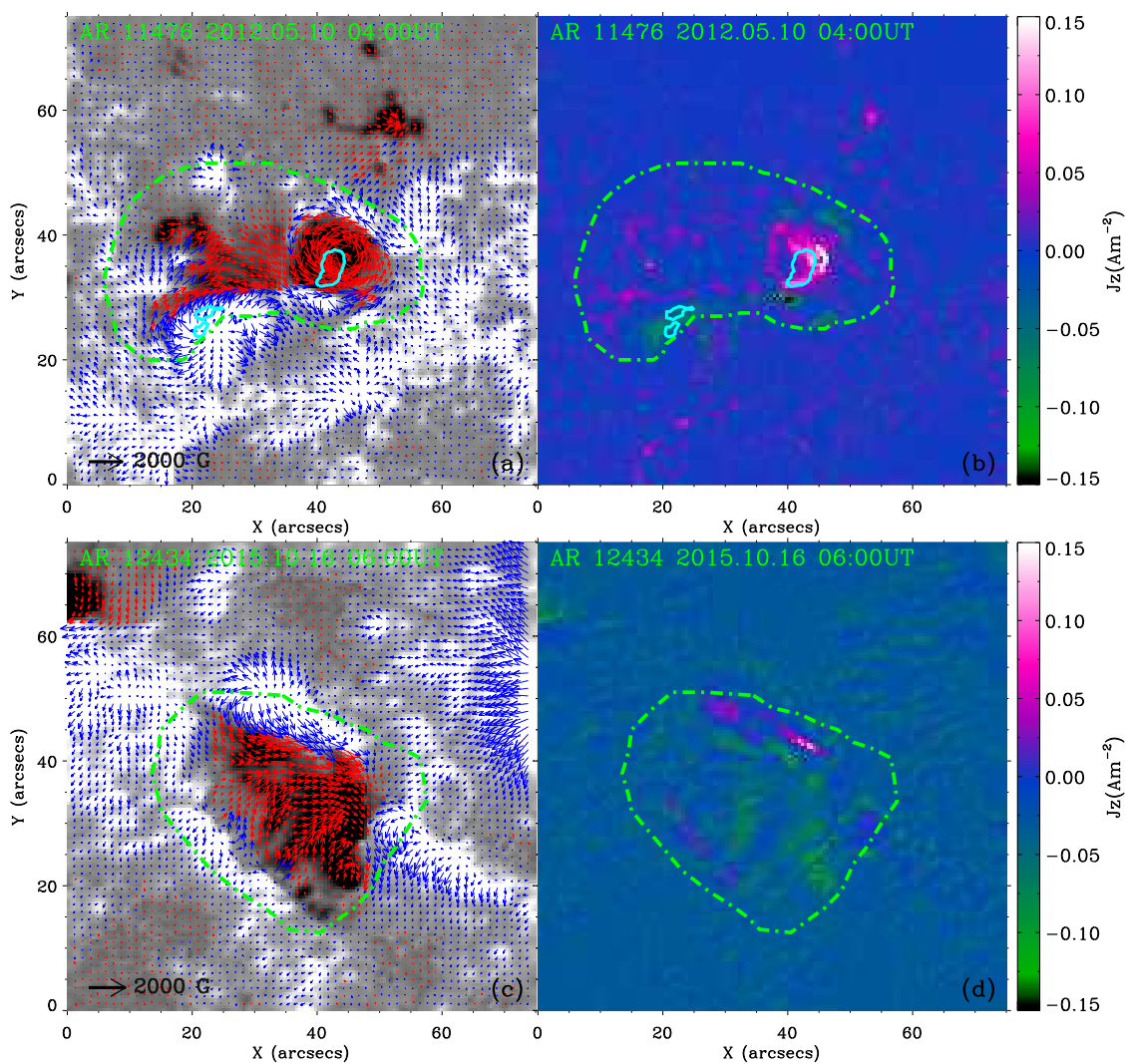}}
\caption{ Spatial distribution of the radial magnetic field and radial electric current ($J_z$) in AR 11476 before an M5.7 flare (upper panels) and in AR 12434 before an M1.1 flare (bottom panels). The green lines mark the circular ribbons. The cyan contours in panels (a) and (b) mark the WL enhancements during the M5.7 flare with a level $(I_p-I_0)/I_0=0.1$. Here $I_p$ and $I_0$ refer to the HMI continuum intensities at 04:16 UT and 04:14 UT, respectively.}
\end{figure}

\begin{figure}[!ht]
\centerline{\includegraphics[width=0.6\textwidth]{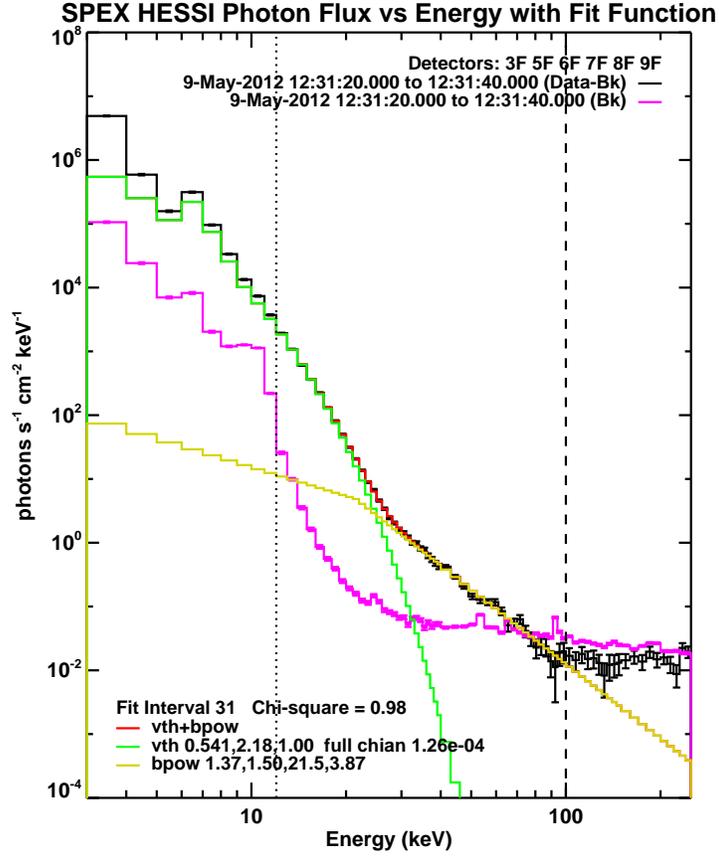}}
\caption{ \textit{RHESSI} energy spectrum around the peak time of an M4.7 flare on 2012 May 9 in AR 11476. The black curve is the observed spectrum after subtracting the background (pink curve), which is fitted by a variable thermal function (green) and a non-thermal broken power law function (yellow). The dotted and dashed vertical lines indicate the fitted energy range. }
\end{figure}

\begin{figure}[!ht]
\centerline{\includegraphics[width=0.9\textwidth]{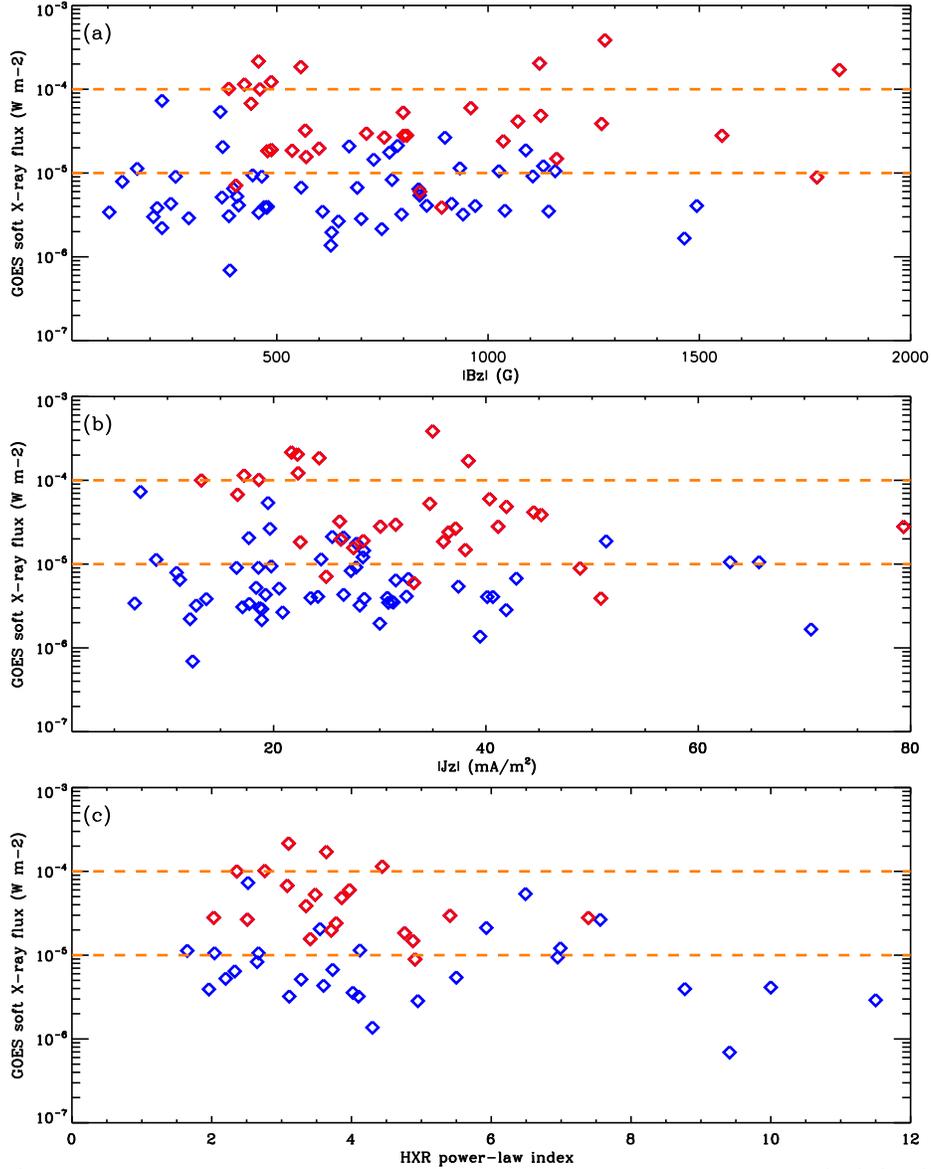}}
\caption{Scatter plots showing the relationship between the peak of \textit{GOES} 1-8 \AA\ X-ray flux and the radial component of the unsigned magnetic field strength ($|Bz|$, a), the electric current ($|J_z|$, b), and the HXR power-law index (c). The red and blue diamonds represent WLFs and non-WL flares, respectively.}
\end{figure}

\begin{figure}[!ht]
\centerline{\includegraphics[width=1\textwidth]{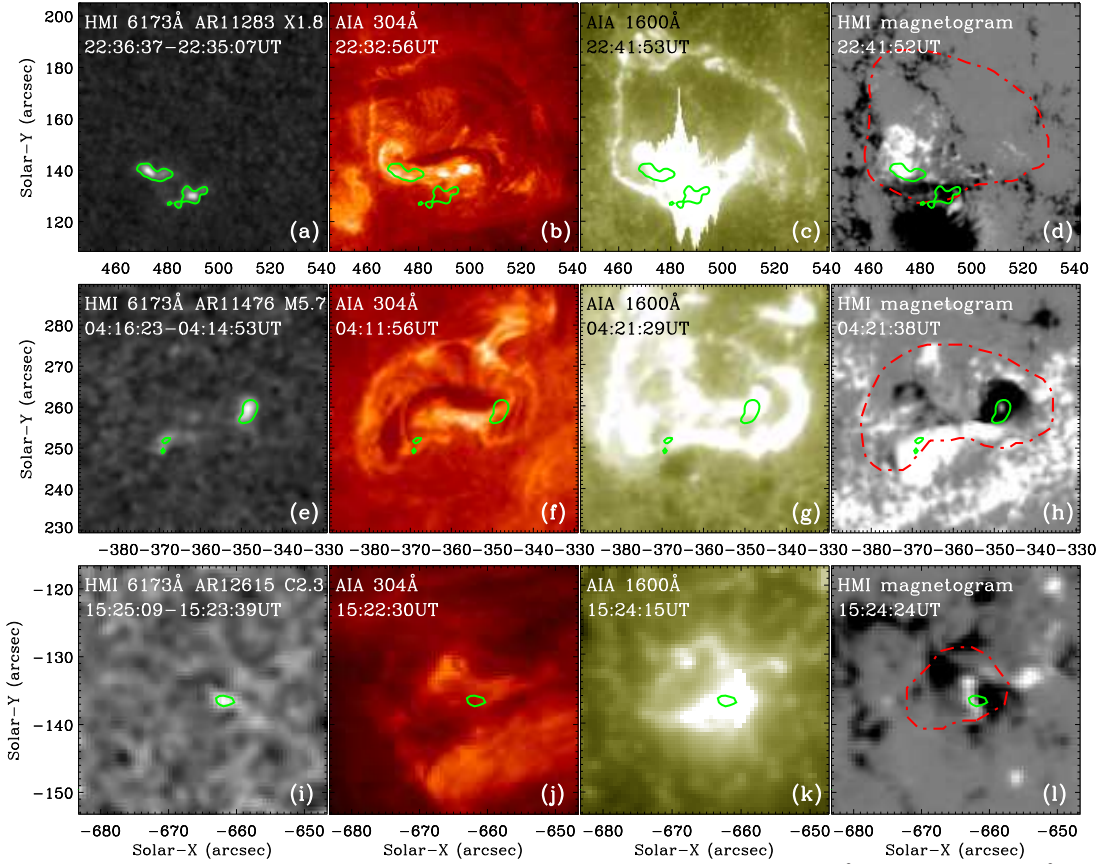}}
\caption{ Three WLFs observed in HMI continuum, AIA 304 \AA\ , AIA 1600 \AA\ and HMI line-of-sight magnetogram. Panels (a)-(d) show the results for an X1.8 flare in NOAA AR 11283; panels (e)-(h) are for an M5.7 flare in NOAA AR 11476; panels (i)-(l) are for a C2.3 flare in NOAA AR 12615. Green contours mark the WL kernels. For the X- and M- class flares, the contour level is $(I_p-I_0)/I_0=0.1$. For the C-class flare, the contour level is $(I_p-I_0)/I_0=0.05$. The red lines mark the circular flare ribbons.}
\end{figure}

\begin{figure}[!ht]
\centerline{\includegraphics[width=1.\textwidth]{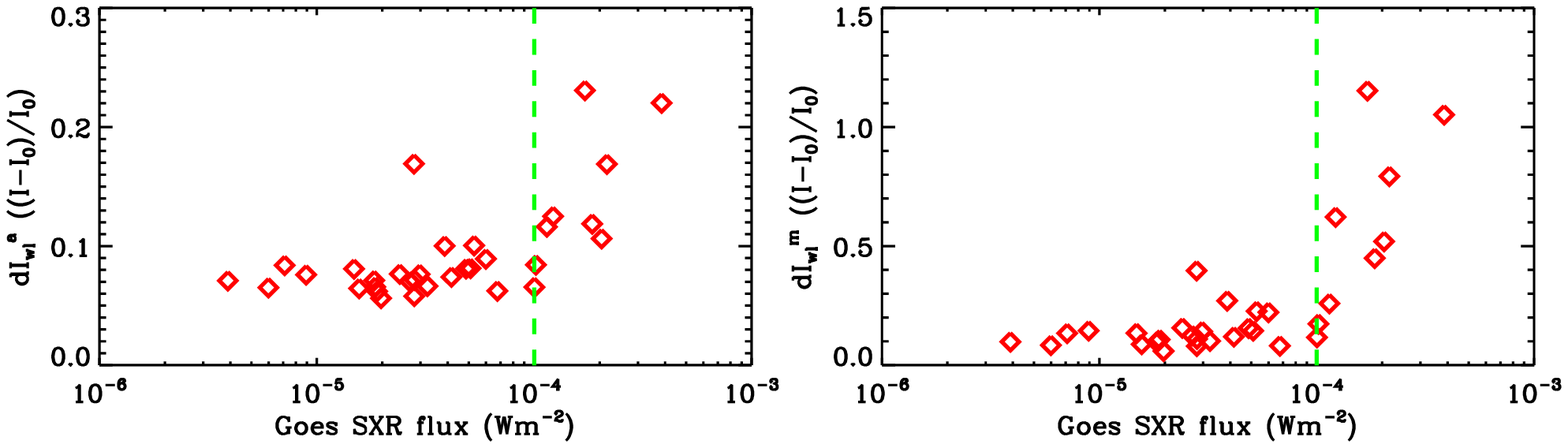}}
\caption{ Relationship between the WL enhancement and the peak of \textit{GOES} 1-8 \AA\ X-ray flux. The average ($dI_{wl}^a$) and maximum ($dI_{wl}^m$) values of WL enhancement are shown on the left and right, respectively. The green dashed line in each panel corresponds to the \textit{GOES} class of X1.0.}
\end{figure}

\begin{figure}[!ht]
\centerline{\includegraphics[width=1.\textwidth]{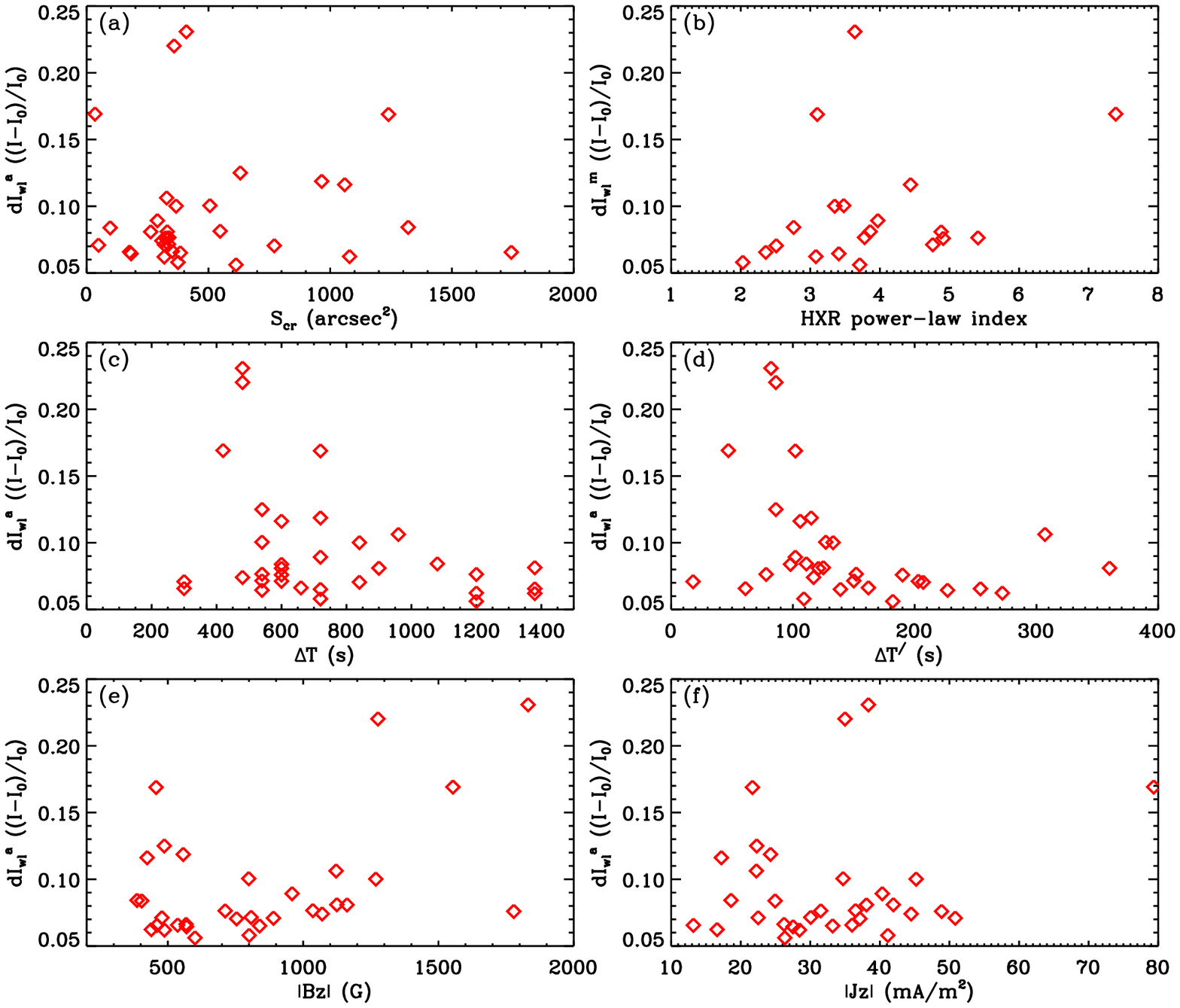}}
\caption{ Scatter plots showing the relationship between WL enhancement and the area enclosed by the circular ribbon (a), HXR power-law index (b), flare duration (c), impulsive phase duration(d), $|B_z|$ (e) and $|J_z|$ (f).}
\end{figure}

\end{document}